\pdfoutput=1 

\documentclass[%
reprint,
groupedaddress,
 amsmath,amssymb,
 aip,
prl,
floatfix,
]{revtex4-1}

\usepackage{xcolor} 

\usepackage{graphicx}
\usepackage{dcolumn}
\usepackage{bm}
\usepackage{hyperref}
\usepackage[mathlines]{lineno}

\begin{document}

\preprint{APS/123-QED}

\title{Demonstration of superconducting micromachined cavities}

\author{T. Brecht}
\email{teresa.brecht@yale.edu}
\author{M. Reagor}
\author{Y. Chu}
\author{W. Pfaff} 
\author{C. Wang}
\author{L. Frunzio}
\author{M.H. Devoret}
\author{R.J. Schoelkopf}
\affiliation{Department of Applied Physics, Yale University, New Haven, Connecticut 06511, USA}

\date{\today}

\begin{abstract}
Superconducting enclosures will be key components of scalable quantum computing devices based on circuit quantum electrodynamics (cQED). Within a densely integrated device, they can protect qubits from noise and serve as quantum memory units. Whether constructed by machining bulk pieces of metal or microfabricating wafers, 3D enclosures are typically assembled from two or more parts. The resulting seams potentially dissipate crossing currents and limit performance. In this Letter, we present measured quality factors of superconducting cavity resonators of several materials, dimensions and seam locations. We observe that superconducting indium can be a low-loss RF conductor and form low-loss seams. Leveraging this, we create a superconducting micromachined resonator with indium that has a quality factor of two million despite a greatly reduced mode volume. Inter-layer coupling to this type of resonator is achieved by an aperture located under a planar transmission line. The described techniques demonstrate a proof-of-principle for multilayer microwave integrated quantum circuits for scalable quantum computing.

\end{abstract}

\maketitle


\section{Introduction}

Quantum information devices based on the circuit quantum electrodynamics (cQED) platform are now reaching a level of complexity that demands careful attention to the challenges of connectorization, addressability, and isolation of constituent components. Solutions will ideally allow for dense packing of components and routing of control and readout circuitry while protecting the coherence of quantum states and maintaining the performance of numerous integrated elements.

The semiconductor industry has already solved similar challenges in the classical domain. An exciting prospect is adapting these known methods of 3D integration and packaging for RF components\cite{KatehiLPB:2001dl} to scale-up quantum circuits.  We envision using micromachining, the bulk etching of silicon wafers, to embed cQED components within a multi-wafer construction (Fig.\ref{seampart}a) containing many superconducting qubits, memories, buses, and amplifiers. Therefore, we have proposed a hardware platform called the multilayer microwave integrated quantum circuit (MMIQC).\cite{Brecht:NPJ, MMIQCpatent} Micromachined enclosures, which have been demonstrated in normal metal RF circuitry\cite{Harle:2003tc, Brown:1998uz}, are highly useful in such a device stack if extended to superconducting coatings. They can suppress cross-talk and radiation loss in circuits, or constitute cavity resonators that can serve as quantum memories. Both applications will require 3D superconducting enclosures that are very low-loss, despite highly-confined electric fields and surface currents.

In this Letter, we show that it is crucial to achieve a superconducting bond between micromachined layers in order to achieve a low-loss enclosure.  Further, we demonstrate that indium can fulfill this objective. We first describe a model that allows us to quantify the dissipation that a seam introduces to a cavity mode. We then present measurements of traditionally machined cavity resonators with several materials, dimensions and seam locations chosen to elucidate the seam as a loss mechanism. These measurements reveal that indium can yield high quality factors ($Q_{int}=8 \times 10^7$) and form a superconducting bond for various types of enclosures. Combining this finding with established fabrication techniques, we have created superconducting micromachined cavities. As we describe, such devices are particularly susceptible to loss at seams, but with indium, they can achieve quality factors of up to two million. They can also be coupled with planar techniques, making them compatible with future multi-wafer stacks containing cQED experiments.

\section{The seam as a loss mechanism in cavity resonators}

 \begin{figure}
\includegraphics[scale = 1,angle=0]{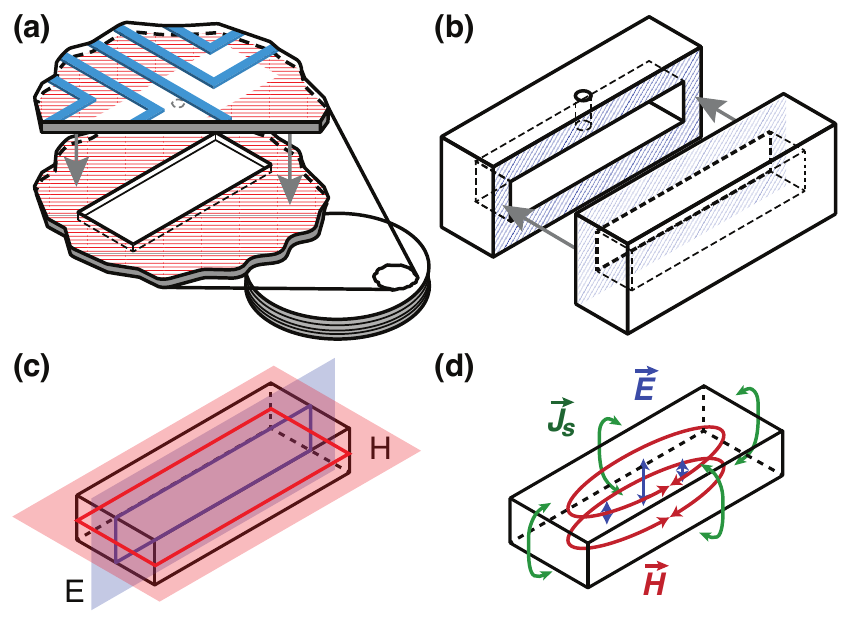}
\caption{
	\textbf{Seam locations in rectanglular cavities.}
    \textbf{(a)} A micromachined cavity formed between two wafers. By coupling to a planar transmission line from an adjacent layer, this unit is suitable for integration in a multilayer microwave integrated circuit. Compared to (b), the height of the cavity is reduced by a factor of 10 and the seam has perpendicular orientation.
    \textbf{(b)} Cavity used in some cQED experiments.\cite{Paik:2011hd} In C(a) and (b), surfaces that meet to form a seam are patterned with red and blue lines.
    \textbf{(c)} Seam locations in the rectangular cavity are shown by cuts in planes H and E.
    \textbf{(d)} Electric (blue) and magnetic (red) field lines of the TE101 mode, and the corresponding surface currents (green).
    }
\label{seampart}
\end{figure}

In addition to dielectric and conductor losses \cite{Zmuidzinas:2012kh,Geerlings:2012ps,Gao:2006fe,Wenner:2011co,Reagor:2013tq}, a cavity assembled from two conductors can experience loss from an imperfect seam.\cite{EBWaccelerators, waveguideflange, PhotonicCrystalJoint, HarocheFBCavity}  Currents crossing the joint between two superconducting walls encounter a discontinuity due to microscopic voids, oxides or other surface impurities.  Therefore, the types of cavities that are commonly used in cQED experiments \cite{Paik:2011hd} are designed to mitigate seam losses (Fig.~\ref{seampart}b). These cavities are assembled from two symmetric halves, split along a plane E shown in Fig.~\ref{seampart}c. The cavity's TE$_{n0m}$ modes ideally have no currents crossing the seam for this configuration (Fig. \ref{seampart}d). Other cavity resonators that have this property include the TE$_{011}$ mode of a cylindrical cavity\cite{Reagor:2013tq, TE011:1968}, coaxial quarter-wave cavities\cite{Reagor:2015}, as well as cavities constructed by extruding methods, such as hydroforming. However, none of these cavity constructions are practical for wafer-processing. MMIQCs' constitutive micromachined enclosures (Fig.~\ref{seampart}a) are etched into silicon wafers and have seams  that experience large currents: those parallel to plane H in Figure~\ref{seampart}c.

We now quantify the dependence of seam losses on the cavity geometry. We model the seam as a distributed port around a path $\vec{l}$ with a total length $L$ and total conductance $G_{seam}$.  The cavity fields are accompanied by surface currents $\vec{J_s}$ that may pass across the seam and dissipate power
\begin{equation}
P_{dis} = \frac{1}{2G_{seam}}
L \int_{seam}|\vec{J_s} \times \hat{l}|^2 dl.
\end{equation}
If it is damped solely by seam losses, a cavity mode of frequency $\omega$ and total energy $E_{tot}$ has a quality factor $Q_i$ given by
\begin{equation}
\frac{1}{Q_i} = \frac{1}{\omega} \frac{P_{dis}}{E_{tot}} =
\frac{1}{G_{seam}}
\left[\frac{L \int_{seam} |\vec{J_s} \times \hat{l}|^2 dl}
{\omega \mu_o \int_{tot} |\vec{H}|^2 dV} \right]
=\frac{y_{seam}}{g_{seam}},
\end{equation}
where the field $\vec{H}$ is integrated over the volume $V$ of the mode and $\mu_o$ is the magnetic permeability.  We identify the expression in square brackets as the admittance, $Y_{seam}$, of the cavity presented to the seam. This admittance is zero when the seam is placed such that there are no perpendicular surface currents.  In order to compare intrinsic seam properties in different cavity constructions, we introduce the conductance per unit length $g_{seam}=G_{seam}/L$ and admittance per unit length $y_{seam}=Y_{seam}/L$. Using this model, we can associate $y_{seam}$ with the seam location and cavity fields and $g_{seam}$ with materials properties in the seam region.

From these equations and the known fields of the TE101 mode (Fig. \ref{seampart}d), we see that plane H of Figure~1c maximizes $y_{seam}$. However, this orientation of seam is the natural consequence of the micromachining process \citep{Brown:1998uz}. Therefore, for a low-loss micromachined cavity, it is imperative to develop a method of wafer-scale superconducting bonding that maximizes $g_{seam}$. By measuring several seam-limited quality factors, we can extract values of $g_{seam}$ for cavities bonded using different materials and techniques.

\section{Measurements of machined superconducting cavities}

\begin{figure}[t]
\begin{center}
\includegraphics[scale = 1,angle=0]{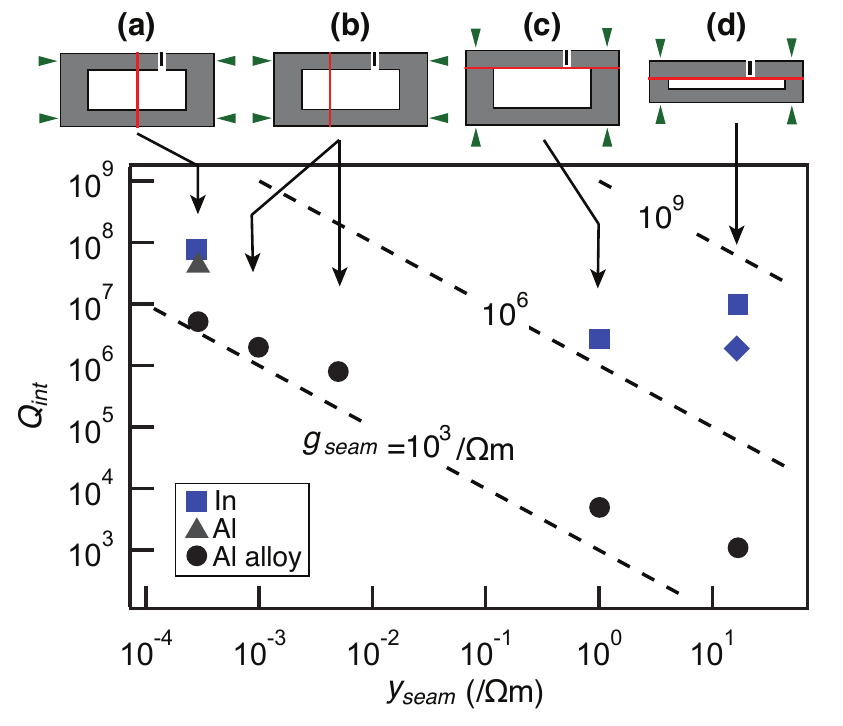}
\caption{
    \textbf{Quality factor measurements of Al and In cavities.}
    Internal quality factors of cavity resonators plotted against seam admittance. Each $y_{seam}$ is calculated using the known fields of the cavity geometry and a seam path. Black dashed lines correspond to several $g_{seam}$ values.  \textbf{(a)-(d)} Pictorial descriptions show a T-plane cross-section of several cavity constructions. Dimensions of cavities of types (a)-(c) were approximately 33$\times$18$\times$5 mm. The dimensions of cavity type (d) were 28$\times$19$\times$0.3 mm and 22$\times$24$\times$0.3 mm. All cavities were coupled using a pin in a sub-cutoff waveguide and had TE101 resonances between 9.45 - 9.54 GHz. Measurements were taken in hanger configuration at temperatures of about 20 mK. With the exception of the diamond point, the two halves of all the cavities were bolted together with four screws that remained during measurement, applying force indicated by green arrows. We plot one In-plated micromachined cavity in Si bonded to In-plated Cu without the use of screws or clamping (blue diamond).
	}
\label{plot}
\end{center}
\end{figure}

Fig. \ref{plot} presents several measurements of quality factors for rectangular waveguide cavity resonators. These devices were machined in bulk metal to various geometries and assembled with several seam locations to study the effect different $y_{seam}$. Moreover, various materials were used: 6061 aluminum alloy (black circles), aluminum of $> 4$N purity (grey triangle), or electroplated indium (blue squares/diamond), to study how material choices contribute to $g_{seam}$. In the following, we describe the cavities in order of increasing seam participation.

We begin with a rectangular cavity geometry similar to those used in 3D cQED experiments\cite{Paik:2011hd}, assembled in two halves that meet at a seam in plane E (Fig. \ref{seampart}b,c, Fig. \ref{plot}a). Machining imprecision of $\pm$ 2 mils may cause deviation from the ideal $y_{seam}=0$.  In bulk Al, these cavities typically have $Q_{int}=10^6-10^7$. Pure ($> 4$N) Al cavities, when chemically etched, have higher $Q_{int}$ due to reduced surface loss. The same cavity geometry was also machined in OFHC Cu and electroplated with In to a thickness of 100 $\mu$m before assembly. The resulting quality factor of $8 \times 10^7$ exceeded that of the Al 4N cavity. Our cavity measurements place a bound on indium's surface quality, described in Ref. \cite{Zmuidzinas:2012kh} as the ratio of a superconductor's surface reactance to surface resistance, of $Q_S > 3 \times 10^3$. 

Next, we describe measurements performed on rectangular cavities with seam participation intentionally increased.  We machined the mating parts in Al alloy to have either a 5\% or 10\% asymmetry along their length such that, once assembled, the E-plane seam was not centered (Fig. \ref{plot}b). The increased $y_{seam}$ is accompanied by 60\% and 85\% reductions in $Q_{int}$ compared to the cavity with symmetric E-plane seam. We also constructed cavities of similar dimensions with seams in the H-plane (Fig. \ref{plot}c). Al cavities with H-plane seams (Fig. \ref{plot}c) had quality factors limited to $5 \times 10^3$. In an identical geometry, we observe that cavities constructed from Cu components electroplated with In had quality factors three orders of magnitude greater ($Q_{int}=2.7 \times 10^6$).

The cavities with the most seam participation, located on the right side of Figure~\ref{plot}, are metal cavities made with a height of 300 $\mu$m, which is feasible to micromachine in a wafer (Fig. \ref{plot}d). In addition to the discussed sensitivity to seam losses, these thin cavities are also more sensitive to surface and conductor losses when compared to larger cavities, with participation ratios scaling inversely with cavity height. It is clear from our measurements, however, that the dominant loss mechanism for these Al cavities is related to the seam.  

We conclude that the In-plated Cu cavities have a $g_{seam}$ far exceeding that of the Al cavities. Taking $g_{seam} = 10^4\,/\Omega \mathrm{m}$ as a likely limit for the Al alloy cavities, this implies an average total conductance of the seams of these cavities of $G_{seam}=1/\left[1.25\,\mu\Omega \right]$.  However, the measured In cavities, with equal $y_{seam}$ posses a higher conductance $g_{seam}=10^8\,/\Omega \mathrm{m}$, meaning that the total conductance of an In seam cavity can be at least as high as $G_{seam}=1/\left[70\,\mathrm{n}\Omega \right]$. We attribute this difference to a comparably weak surface oxide. Another factor may be indium's higher ductility, which is maintained at cryogenic temperatures.\cite{Indium:Reed:1988} In addition to enhanced seam quality, these measurements also indicate that thick In film has a high surface quality.

\begin{figure}
\includegraphics[scale = 1,angle=0]{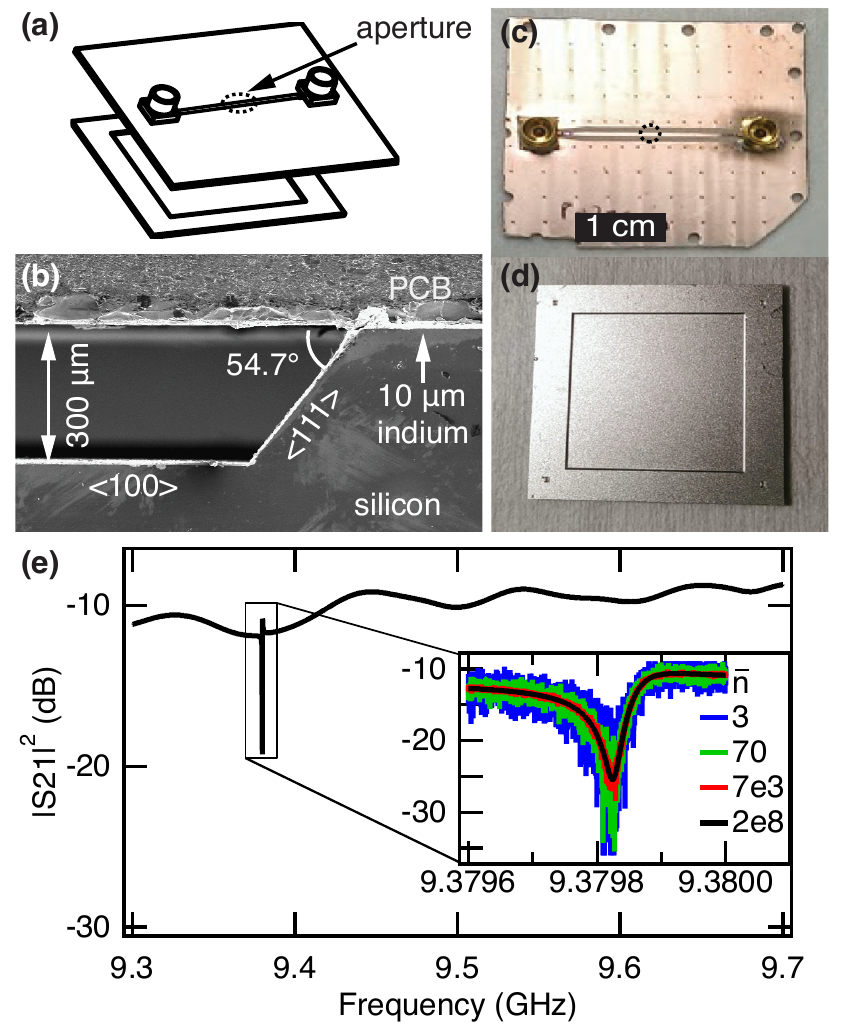}
\caption{
	\textbf{Superconducting micromachined cavity.}  
	\textbf{(a)} Sketch of the construction. 
	\textbf{(b)} SEM of an E-plane cross-section of the device.
	\textbf{(c)} The top chip is made in TMM10i PCB and contains Cu plated vias and 10 $\mu$m In is electroplated on the back side.  An aperture (dashed circle, r=675 $\mu$m) is omitted in the back side metalization to permit coupling of cavity fields through the dielectric to the CPW trace above. 
    \textbf{(d)} The bottom chip consists of a 22 x 24 mm rectangular pit wet etched in a 1 mm thick Si wafer and coated in 100 nm evaporated Cu before electroplating with 10 $\mu$m In.
    \textbf{(e)} A 500 MHz range of the transmission spectrum of the device shows the TE101 resonance.  Measured at 20 mK and $7 \times 10^3$ photons, a fit to the asymmetric lineshape\cite{Khalil:2012jr} yields $Q_{int}=4.6 \times 10^5$ and $Q_c=1.4 \times 10^5$.}
     

\label{umachine}
\end{figure}

\section{Measurements of superconducting indium micromachined cavities}

Having explored the properties of electroplated indium as a superconducting bonding material using metallic prototypes, we now turn to cavities fabricated in wafers. Compatible with MMIQCs, silicon wafers are etched to form recesses which are then electroplated with indium to form cavities. Details of the fabrication process are described in Ref. \cite{SUPPLEMENT}. 

For initial testing, an In-plated Si chip with etched cavity pit was bonded to a block of In-plated Cu containing a sub-cutoff waveguide for pin coupling in a manner identical to the 3D cavities machined in bulk metal. This bond provided both the mechanical and electrical integrity of the micromachined cavity construction, and can result in a $Q_{int}$ exceeding $10^6$ (Fig. \ref{plot}, blue diamond).

A coupling scheme that is more compatible with a multilayer architecture is achieved using a coplanar waveguide or microstrip to access the fields of the resonator through an aperture in the cavity wall, as described in Refs. \cite{Harle:2003tc,Brown:1998uz,SUPPLEMENT}. In Fig. \ref{umachine}, we present a superconducting micomachined cavity with planar coupling, consisting of an etched cavity mated with a printed circuit board (PCB) top chip. Of several devices with various designed couplings\cite{SUPPLEMENT}, we report a best $Q_{int}$ of $4.6 \times 10^5$ ($Q_c=1.4 \times 10^5$). 

Though it exceeds what could be achieved for an Al coating, this result is inconsistent with quality factors shown to be possible in this geometry, and we attribute it to seam loss. We suspect that there is both an oxide barrier and voids in the seam due to imperfect indium electroplating and a non-optimized bonding procedure. Furthermore, we suspect that the integrity of the bonds suffers from a mismatch in the thermal contraction of different materials that we do not expect in future multilayer Si devices.

\section{Outlook}

From this investigation, we conclude that while micromachined cavities are sensitive to seam losses, high-quality superconducting bonds are possible with the proper selection and treatment of materials. Our results suggest that indium is a low loss ($Q_S>3 \times 10^3$) and viable superconducting bonding material for a multilayer cQED hardware platform. This bond quality ($g_{seam}=10^8\,/\Omega \mathrm{m}$) could likely be improved significantly by incorporating techniques such as surface passivation, heating, and precise alignment that are standard in industrial bump bonding\cite{Inbump2,InBump4}. The RF loss properties of superconducting interconnects and wafer-scale bonds produced with these techniques \cite{Miller:2012vw} is a topic of current investigation.

Micromachining cavity resonators with the process described here offers the advantages of lithographic precision and ease of integration with planar circuitry. We have demonstrated planar fabrication of inter-layer coupling to a micromachined cavity. Similar RF interconnects can mediate couplings in an indium bonded stack of silicon wafers with multiple resonators and other quasi-3D and planar elements.\cite{KatehiLPB:2001dl} The micromachined device presented here is an important step towards developing fabrication strategies necessary for scalable quantum computing.\cite{Brecht:NPJ}

\section{Acknowledgments}

We thank Harvey Moseley and Hanhee Paik for useful conversations, Jan Schroers, and Emily Kinser for assistance with wafer bonding, and Christopher Axline, Philip Reinhold and Michael Hatridge for experimental assistance. This research was supported by the U.S. Army Research Office (W911NF-14-1-0011). All statements of fact, opinion or conclusions contained herein are those of the authors and should not be construed as representing the official views or policies of the U.S. Government. W.P. was supported by NSF grant PHY1309996 and by a fellowship instituted with a Max Planck Research Award from the Alexander von Humboldt Foundation. The Yale Center for Microelectronic Materials and Structures facilities were used for some parts of the fabrication.



\begin{thebibliography}{24}%
\makeatletter
\providecommand \@ifxundefined [1]{%
 \@ifx{#1\undefined}
}%
\providecommand \@ifnum [1]{%
 \ifnum #1\expandafter \@firstoftwo
 \else \expandafter \@secondoftwo
 \fi
}%
\providecommand \@ifx [1]{%
 \ifx #1\expandafter \@firstoftwo
 \else \expandafter \@secondoftwo
 \fi
}%
\providecommand \natexlab [1]{#1}%
\providecommand \enquote  [1]{``#1''}%
\providecommand \bibnamefont  [1]{#1}%
\providecommand \bibfnamefont [1]{#1}%
\providecommand \citenamefont [1]{#1}%
\providecommand \href@noop [0]{\@secondoftwo}%
\providecommand \href [0]{\begingroup \@sanitize@url \@href}%
\providecommand \@href[1]{\@@startlink{#1}\@@href}%
\providecommand \@@href[1]{\endgroup#1\@@endlink}%
\providecommand \@sanitize@url [0]{\catcode `\\12\catcode `\$12\catcode
  `\&12\catcode `\#12\catcode `\^12\catcode `\_12\catcode `\%12\relax}%
\providecommand \@@startlink[1]{}%
\providecommand \@@endlink[0]{}%
\providecommand \url  [0]{\begingroup\@sanitize@url \@url }%
\providecommand \@url [1]{\endgroup\@href {#1}{\urlprefix }}%
\providecommand \urlprefix  [0]{URL }%
\providecommand \Eprint [0]{\href }%
\providecommand \doibase [0]{http://dx.doi.org/}%
\providecommand \selectlanguage [0]{\@gobble}%
\providecommand \bibinfo  [0]{\@secondoftwo}%
\providecommand \bibfield  [0]{\@secondoftwo}%
\providecommand \translation [1]{[#1]}%
\providecommand \BibitemOpen [0]{}%
\providecommand \bibitemStop [0]{}%
\providecommand \bibitemNoStop [0]{.\EOS\space}%
\providecommand \EOS [0]{\spacefactor3000\relax}%
\providecommand \BibitemShut  [1]{\csname bibitem#1\endcsname}%
\let\auto@bib@innerbib\@empty
\bibitem [{\citenamefont {{Katehi, L.P.B}}, \citenamefont {Harvey},\ and\
  \citenamefont {Herrick}(2001)}]{KatehiLPB:2001dl}%
  \BibitemOpen
  \bibfield  {author} {\bibinfo {author} {\bibnamefont {{Katehi, L.P.B}}},
  \bibinfo {author} {\bibfnamefont {J.~F.}\ \bibnamefont {Harvey}}, \ and\
  \bibinfo {author} {\bibfnamefont {K.~J.}\ \bibnamefont {Herrick}},\ }\href
  {\doibase 10.1109/6668.918260} {\bibfield  {journal} {\bibinfo  {journal}
  {IEEE Microw. Mag.}\ }\textbf {\bibinfo {volume} {2}},\ \bibinfo {pages} {30}
  (\bibinfo {year} {2001})}\BibitemShut {NoStop}%
\bibitem [{\citenamefont {Brecht}\ \emph {et~al.}(2015)\citenamefont {Brecht},
  \citenamefont {Pfaff}, \citenamefont {Chu}, \citenamefont {Frunzio},
  \citenamefont {Devoret},\ and\ \citenamefont {Schoelkopf}}]{Brecht:NPJ}%
  \BibitemOpen
  \bibfield  {author} {\bibinfo {author} {\bibfnamefont {T.}~\bibnamefont
  {Brecht}}, \bibinfo {author} {\bibfnamefont {W.}~\bibnamefont {Pfaff}},
  \bibinfo {author} {\bibfnamefont {Y.}~\bibnamefont {Chu}}, \bibinfo {author}
  {\bibfnamefont {L.}~\bibnamefont {Frunzio}}, \bibinfo {author} {\bibfnamefont
  {M.~H.}\ \bibnamefont {Devoret}}, \ and\ \bibinfo {author} {\bibfnamefont
  {R.~J.}\ \bibnamefont {Schoelkopf}},\ }\href@noop {} {\bibfield  {journal}
  {\bibinfo  {journal} {arXiv.org}\ } (\bibinfo {year} {2015})},\ \Eprint
  {http://arxiv.org/abs/1509.01127} {1509.01127} \BibitemShut {NoStop}%
\bibitem [{\citenamefont {Schoelkopf}\ \emph {et~al.}()\citenamefont
  {Schoelkopf}, \citenamefont {Brecht}, \citenamefont {Frunzio},\ and\
  \citenamefont {Devoret}}]{MMIQCpatent}%
  \BibitemOpen
  \bibfield  {author} {\bibinfo {author} {\bibfnamefont {R.~J.}\ \bibnamefont
  {Schoelkopf}}, \bibinfo {author} {\bibfnamefont {T.}~\bibnamefont {Brecht}},
  \bibinfo {author} {\bibfnamefont {L.}~\bibnamefont {Frunzio}}, \ and\
  \bibinfo {author} {\bibfnamefont {M.~H.}~\bibnamefont {Devoret}},\ }\href@noop
  {} {}\bibinfo {note} {WO Patent App. PCT/US2014/012,073}\BibitemShut
  {NoStop}%
\bibitem [{\citenamefont {Harle}(2003)}]{Harle:2003tc}%
  \BibitemOpen
  \bibfield  {author} {\bibinfo {author} {\bibfnamefont {L.}~\bibnamefont
  {Harle}},\ }\emph {\bibinfo {title} {{Microwave micromachined cavity
  filters}}},\ \href@noop {} {Ph.D. thesis},\ \bibinfo  {school} {University of
  Michigan} (\bibinfo {year} {2003})\BibitemShut {NoStop}%
\bibitem [{\citenamefont {Brown}, \citenamefont {Blondy},\ and\ \citenamefont
  {Rebeiz}(1998)}]{Brown:1998uz}%
  \BibitemOpen
  \bibfield  {author} {\bibinfo {author} {\bibfnamefont {A.~R.}\ \bibnamefont
  {Brown}}, \bibinfo {author} {\bibfnamefont {P.}~\bibnamefont {Blondy}}, \
  and\ \bibinfo {author} {\bibfnamefont {G.~M.}\ \bibnamefont {Rebeiz}},\
  }\href {\doibase
  10.1002/(SICI)1099-047X(199907)9:4<326::AID-MMCE4>3.0.CO;2-Y} {\bibfield
  {journal} {\bibinfo  {journal} {Int. J. RF Microw. Comp. Aid. Eng.}\ }\textbf
  {\bibinfo {volume} {9}},\ \bibinfo {pages} {1} (\bibinfo {year}
  {1998})}\BibitemShut {NoStop}%
\bibitem [{\citenamefont {Paik}\ \emph {et~al.}(2011)\citenamefont {Paik},
  \citenamefont {Schuster}, \citenamefont {Bishop}, \citenamefont {Kirchmair},
  \citenamefont {Catelani}, \citenamefont {Sears}, \citenamefont {Johnson},
  \citenamefont {Reagor}, \citenamefont {Frunzio}, \citenamefont {Glazman},
  \citenamefont {Girvin}, \citenamefont {Devoret},\ and\ \citenamefont
  {Schoelkopf}}]{Paik:2011hd}%
  \BibitemOpen
  \bibfield  {author} {\bibinfo {author} {\bibfnamefont {H.}~\bibnamefont
  {Paik}}, \bibinfo {author} {\bibfnamefont {D.~I.}\ \bibnamefont {Schuster}},
  \bibinfo {author} {\bibfnamefont {L.~S.}\ \bibnamefont {Bishop}}, \bibinfo
  {author} {\bibfnamefont {G.}~\bibnamefont {Kirchmair}}, \bibinfo {author}
  {\bibfnamefont {G.}~\bibnamefont {Catelani}}, \bibinfo {author}
  {\bibfnamefont {A.~P.}\ \bibnamefont {Sears}}, \bibinfo {author}
  {\bibfnamefont {B.~R.}\ \bibnamefont {Johnson}}, \bibinfo {author}
  {\bibfnamefont {M.~J.}\ \bibnamefont {Reagor}}, \bibinfo {author}
  {\bibfnamefont {L.}~\bibnamefont {Frunzio}}, \bibinfo {author} {\bibfnamefont
  {L.~I.}\ \bibnamefont {Glazman}}, \bibinfo {author} {\bibfnamefont {S.~M.}\
  \bibnamefont {Girvin}}, \bibinfo {author} {\bibfnamefont {M.~H.}\
  \bibnamefont {Devoret}}, \ and\ \bibinfo {author} {\bibfnamefont {R.~J.}\
  \bibnamefont {Schoelkopf}},\ }\href {\doibase 10.1103/PhysRevLett.107.240501}
  {\bibfield  {journal} {\bibinfo  {journal} {Phys. Rev. Lett.}\ }\textbf
  {\bibinfo {volume} {107}},\ \bibinfo {pages} {240501} (\bibinfo {year}
  {2011})}\BibitemShut {NoStop}%
\bibitem [{\citenamefont {Zmuidzinas}(2012)}]{Zmuidzinas:2012kh}%
  \BibitemOpen
  \bibfield  {author} {\bibinfo {author} {\bibfnamefont {J.}~\bibnamefont
  {Zmuidzinas}},\ }\href {\doibase 10.1146/annurev-conmatphys-020911-125022}
  {\bibfield  {journal} {\bibinfo  {journal} {Ann. Rev. Cond. Matter Phys.}\
  }\textbf {\bibinfo {volume} {3}},\ \bibinfo {pages} {169} (\bibinfo {year}
  {2012})}\BibitemShut {NoStop}%
\bibitem [{\citenamefont {Geerlings}\ \emph {et~al.}(2012)\citenamefont
  {Geerlings}, \citenamefont {Shankar}, \citenamefont {Edwards}, \citenamefont
  {Frunzio}, \citenamefont {Schoelkopf},\ and\ \citenamefont
  {Devoret}}]{Geerlings:2012ps}%
  \BibitemOpen
  \bibfield  {author} {\bibinfo {author} {\bibfnamefont {K.}~\bibnamefont
  {Geerlings}}, \bibinfo {author} {\bibfnamefont {S.}~\bibnamefont {Shankar}},
  \bibinfo {author} {\bibfnamefont {E.}~\bibnamefont {Edwards}}, \bibinfo
  {author} {\bibfnamefont {L.}~\bibnamefont {Frunzio}}, \bibinfo {author}
  {\bibfnamefont {R.}~\bibnamefont {Schoelkopf}}, \ and\ \bibinfo {author}
  {\bibfnamefont {M.}~\bibnamefont {Devoret}},\ }\href {\doibase
  10.1063/1.4710520} {\bibfield  {journal} {\bibinfo  {journal} {Appl. Phys.
  Lett.}\ }\textbf {\bibinfo {volume} {100}},\ \bibinfo {pages} {192601}
  (\bibinfo {year} {2012})}\BibitemShut {NoStop}%
\bibitem [{\citenamefont {Gao}\ \emph {et~al.}(2006)\citenamefont {Gao},
  \citenamefont {Zmuidzinas}, \citenamefont {Mazin}, \citenamefont {Day},\ and\
  \citenamefont {Leduc}}]{Gao:2006fe}%
  \BibitemOpen
  \bibfield  {author} {\bibinfo {author} {\bibfnamefont {J.}~\bibnamefont
  {Gao}}, \bibinfo {author} {\bibfnamefont {J.}~\bibnamefont {Zmuidzinas}},
  \bibinfo {author} {\bibfnamefont {B.~A.}\ \bibnamefont {Mazin}}, \bibinfo
  {author} {\bibfnamefont {P.~K.}\ \bibnamefont {Day}}, \ and\ \bibinfo
  {author} {\bibfnamefont {H.~G.}\ \bibnamefont {Leduc}},\ }\href {\doibase
  10.1016/j.nima.2005.12.075} {\bibfield  {journal} {\bibinfo  {journal} {Nucl.
  Inst. Meth. A}\ }\textbf {\bibinfo {volume} {559}},\ \bibinfo {pages} {585}
  (\bibinfo {year} {2006})}\BibitemShut {NoStop}%
\bibitem [{\citenamefont {Wenner}\ \emph {et~al.}(2011)\citenamefont {Wenner},
  \citenamefont {Barends}, \citenamefont {Bialczak}, \citenamefont {Chen},
  \citenamefont {Kelly}, \citenamefont {Lucero}, \citenamefont {Mariantoni},
  \citenamefont {Megrant}, \citenamefont {O'Malley}, \citenamefont {Sank},
  \citenamefont {Vainsencher}, \citenamefont {Wang}, \citenamefont {White},
  \citenamefont {Yin}, \citenamefont {Zhao}, \citenamefont {Cleland},\ and\
  \citenamefont {Martinis}}]{Wenner:2011co}%
  \BibitemOpen
  \bibfield  {author} {\bibinfo {author} {\bibfnamefont {J.}~\bibnamefont
  {Wenner}}, \bibinfo {author} {\bibfnamefont {R.}~\bibnamefont {Barends}},
  \bibinfo {author} {\bibfnamefont {R.~C.}\ \bibnamefont {Bialczak}}, \bibinfo
  {author} {\bibfnamefont {Y.}~\bibnamefont {Chen}}, \bibinfo {author}
  {\bibfnamefont {J.}~\bibnamefont {Kelly}}, \bibinfo {author} {\bibfnamefont
  {E.}~\bibnamefont {Lucero}}, \bibinfo {author} {\bibfnamefont
  {M.}~\bibnamefont {Mariantoni}}, \bibinfo {author} {\bibfnamefont
  {A.}~\bibnamefont {Megrant}}, \bibinfo {author} {\bibfnamefont {P.~J.~J.}\
  \bibnamefont {O'Malley}}, \bibinfo {author} {\bibfnamefont {D.}~\bibnamefont
  {Sank}}, \bibinfo {author} {\bibfnamefont {A.}~\bibnamefont {Vainsencher}},
  \bibinfo {author} {\bibfnamefont {H.}~\bibnamefont {Wang}}, \bibinfo {author}
  {\bibfnamefont {T.~C.}\ \bibnamefont {White}}, \bibinfo {author}
  {\bibfnamefont {Y.}~\bibnamefont {Yin}}, \bibinfo {author} {\bibfnamefont
  {J.}~\bibnamefont {Zhao}}, \bibinfo {author} {\bibfnamefont {A.~N.}\
  \bibnamefont {Cleland}}, \ and\ \bibinfo {author} {\bibfnamefont {J.~M.}\
  \bibnamefont {Martinis}},\ }\href {\doibase 10.1063/1.3637047} {\bibfield
  {journal} {\bibinfo  {journal} {Appl. Phys. Lett.}\ }\textbf {\bibinfo
  {volume} {99}},\ \bibinfo {pages} {113513} (\bibinfo {year}
  {2011})}\BibitemShut {NoStop}%
\bibitem [{\citenamefont {Reagor}\ \emph {et~al.}(2013)\citenamefont {Reagor},
  \citenamefont {Paik}, \citenamefont {Catelani}, \citenamefont {Sun},
  \citenamefont {Axline}, \citenamefont {Holland}, \citenamefont {Pop},
  \citenamefont {Masluk}, \citenamefont {Brecht}, \citenamefont {Frunzio},
  \citenamefont {Devoret}, \citenamefont {Glazman},\ and\ \citenamefont
  {Schoelkopf}}]{Reagor:2013tq}%
  \BibitemOpen
  \bibfield  {author} {\bibinfo {author} {\bibfnamefont {M.}~\bibnamefont
  {Reagor}}, \bibinfo {author} {\bibfnamefont {H.}~\bibnamefont {Paik}},
  \bibinfo {author} {\bibfnamefont {G.}~\bibnamefont {Catelani}}, \bibinfo
  {author} {\bibfnamefont {L.}~\bibnamefont {Sun}}, \bibinfo {author}
  {\bibfnamefont {C.}~\bibnamefont {Axline}}, \bibinfo {author} {\bibfnamefont
  {E.}~\bibnamefont {Holland}}, \bibinfo {author} {\bibfnamefont {I.~M.}\
  \bibnamefont {Pop}}, \bibinfo {author} {\bibfnamefont {N.~A.}\ \bibnamefont
  {Masluk}}, \bibinfo {author} {\bibfnamefont {T.}~\bibnamefont {Brecht}},
  \bibinfo {author} {\bibfnamefont {L.}~\bibnamefont {Frunzio}}, \bibinfo
  {author} {\bibfnamefont {M.~H.}\ \bibnamefont {Devoret}}, \bibinfo {author}
  {\bibfnamefont {L.}~\bibnamefont {Glazman}}, \ and\ \bibinfo {author}
  {\bibfnamefont {R.~J.}\ \bibnamefont {Schoelkopf}},\ }\href {\doibase
  10.1063/1.4807015} {\bibfield  {journal} {\bibinfo  {journal} {Appl. Phys.
  Lett.}\ }\textbf {\bibinfo {volume} {102}},\ \bibinfo {pages} {192604}
  (\bibinfo {year} {2013})}\BibitemShut {NoStop}%
\bibitem [{\citenamefont {Singer}\ \emph {et~al.}(2011)\citenamefont {Singer},
  \citenamefont {Singer}, \citenamefont {Aderhold}, \citenamefont {Ermakov},
  \citenamefont {Twarowski}, \citenamefont {Crooks}, \citenamefont {Hoss},
  \citenamefont {Sch\"olz},\ and\ \citenamefont {Spaniol}}]{EBWaccelerators}%
  \BibitemOpen
  \bibfield  {author} {\bibinfo {author} {\bibfnamefont {W.}~\bibnamefont
  {Singer}}, \bibinfo {author} {\bibfnamefont {X.}~\bibnamefont {Singer}},
  \bibinfo {author} {\bibfnamefont {S.}~\bibnamefont {Aderhold}}, \bibinfo
  {author} {\bibfnamefont {A.}~\bibnamefont {Ermakov}}, \bibinfo {author}
  {\bibfnamefont {K.}~\bibnamefont {Twarowski}}, \bibinfo {author}
  {\bibfnamefont {R.}~\bibnamefont {Crooks}}, \bibinfo {author} {\bibfnamefont
  {M.}~\bibnamefont {Hoss}}, \bibinfo {author} {\bibfnamefont {F.}~\bibnamefont
  {Sch\"olz}}, \ and\ \bibinfo {author} {\bibfnamefont {B.}~\bibnamefont
  {Spaniol}},\ }\href {\doibase 10.1103/PhysRevSTAB.14.050702} {\bibfield
  {journal} {\bibinfo  {journal} {Phys. Rev. ST Accel. Beams}\ }\textbf
  {\bibinfo {volume} {14}},\ \bibinfo {pages} {050702} (\bibinfo {year}
  {2011})}\BibitemShut {NoStop}%
\bibitem [{\citenamefont {Harvey}(1955)}]{waveguideflange}%
  \BibitemOpen
  \bibfield  {author} {\bibinfo {author} {\bibfnamefont {A.}~\bibnamefont
  {Harvey}},\ }\href {\doibase 10.1049/pi-b-1.1955.0095} {\bibfield  {journal}
  {\bibinfo  {journal} {Proc. IEE - Part B: Radio and Electronic Engineering}\
  }\textbf {\bibinfo {volume} {102}},\ \bibinfo {pages} {493} (\bibinfo {year}
  {1955})}\BibitemShut {NoStop}%
\bibitem [{\citenamefont {Hesler}()}]{PhotonicCrystalJoint}%
  \BibitemOpen
  \bibfield  {author} {\bibinfo {author} {\bibfnamefont {J.}~\bibnamefont
  {Hesler}},\ }in\ \href {\doibase 10.1109/MWSYM.2001.967009} {\emph {\bibinfo
  {booktitle} {Microwave Symposium Digest, 2001 IEEE MTT-S International}}},\
  \bibinfo {series} {Phoenix, AZ, USA, 20-24 May 2001}, Vol.~\bibinfo {volume}
  {2}\ (\bibinfo  {publisher} {IEEE, 2001})\ p.\ \bibinfo {pages}
  {783}\BibitemShut {NoStop}%
\bibitem [{\citenamefont {Kuhr}\ \emph {et~al.}(2007)\citenamefont {Kuhr},
  \citenamefont {Gleyzes}, \citenamefont {Guerlin}, \citenamefont {Bernu},
  \citenamefont {Hoff}, \citenamefont {Deléglise}, \citenamefont {Osnaghi},
  \citenamefont {Brune}, \citenamefont {Raimond}, \citenamefont {Haroche},
  \citenamefont {Jacques}, \citenamefont {Bosland},\ and\ \citenamefont
  {Visentin}}]{HarocheFBCavity}%
  \BibitemOpen
  \bibfield  {author} {\bibinfo {author} {\bibfnamefont {S.}~\bibnamefont
  {Kuhr}}, \bibinfo {author} {\bibfnamefont {S.}~\bibnamefont {Gleyzes}},
  \bibinfo {author} {\bibfnamefont {C.}~\bibnamefont {Guerlin}}, \bibinfo
  {author} {\bibfnamefont {J.}~\bibnamefont {Bernu}}, \bibinfo {author}
  {\bibfnamefont {U.~B.}\ \bibnamefont {Hoff}}, \bibinfo {author}
  {\bibfnamefont {S.}~\bibnamefont {Deléglise}}, \bibinfo {author}
  {\bibfnamefont {S.}~\bibnamefont {Osnaghi}}, \bibinfo {author} {\bibfnamefont
  {M.}~\bibnamefont {Brune}}, \bibinfo {author} {\bibfnamefont {J.-M.}\
  \bibnamefont {Raimond}}, \bibinfo {author} {\bibfnamefont {S.}~\bibnamefont
  {Haroche}}, \bibinfo {author} {\bibfnamefont {E.}~\bibnamefont {Jacques}},
  \bibinfo {author} {\bibfnamefont {P.}~\bibnamefont {Bosland}}, \ and\
  \bibinfo {author} {\bibfnamefont {B.}~\bibnamefont {Visentin}},\ }\href
  {\doibase 10.1063/1.2724816} {\bibfield  {journal} {\bibinfo  {journal}
  {Appl. Phys. Lett.}\ }\textbf {\bibinfo {volume} {90}},\ \bibinfo {eid}
  {164101} (\bibinfo {year} {2007})}\BibitemShut {NoStop}%
\bibitem [{\citenamefont {Turneaure}\ and\ \citenamefont
  {Weissman}(1968)}]{TE011:1968}%
  \BibitemOpen
  \bibfield  {author} {\bibinfo {author} {\bibfnamefont {J.~P.}\ \bibnamefont
  {Turneaure}}\ and\ \bibinfo {author} {\bibfnamefont {I.}~\bibnamefont
  {Weissman}},\ }\href {\doibase 10.1063/1.1656986} {\bibfield  {journal}
  {\bibinfo  {journal} {J. Appl. Phys.}\ }\textbf {\bibinfo {volume} {39}},\
  \bibinfo {pages} {4417} (\bibinfo {year} {1968})}\BibitemShut {NoStop}%
\bibitem [{\citenamefont {Reagor}\ \emph {et~al.}(2015)\citenamefont {Reagor},
  \citenamefont {Pfaff}, \citenamefont {Axline}, \citenamefont {Heeres},
  \citenamefont {Ofek}, \citenamefont {Sliwa}, \citenamefont {Holland},
  \citenamefont {Wang}, \citenamefont {Blumoff}, \citenamefont {Chou},
  \citenamefont {Hatridge}, \citenamefont {Frunzio}, \citenamefont {Devoret},\
  and\ \citenamefont {Schoelkopf}}]{Reagor:2015}%
  \BibitemOpen
  \bibfield  {author} {\bibinfo {author} {\bibfnamefont {M.}~\bibnamefont
  {Reagor}}, \bibinfo {author} {\bibfnamefont {W.}~\bibnamefont {Pfaff}},
  \bibinfo {author} {\bibfnamefont {C.}~\bibnamefont {Axline}}, \bibinfo
  {author} {\bibfnamefont {R.}~\bibnamefont {Heeres}}, \bibinfo {author}
  {\bibfnamefont {N.}~\bibnamefont {Ofek}}, \bibinfo {author} {\bibfnamefont
  {K.}~\bibnamefont {Sliwa}}, \bibinfo {author} {\bibfnamefont
  {E.}~\bibnamefont {Holland}}, \bibinfo {author} {\bibfnamefont
  {C.}~\bibnamefont {Wang}}, \bibinfo {author} {\bibfnamefont {J.}~\bibnamefont
  {Blumoff}}, \bibinfo {author} {\bibfnamefont {K.}~\bibnamefont {Chou}},
  \bibinfo {author} {\bibfnamefont {M.~J.}\ \bibnamefont {Hatridge}}, \bibinfo
  {author} {\bibfnamefont {L.}~\bibnamefont {Frunzio}}, \bibinfo {author}
  {\bibfnamefont {M.~H.}\ \bibnamefont {Devoret}}, \ and\ \bibinfo {author}
  {\bibfnamefont {R.~J.}\ \bibnamefont {Schoelkopf}},\ }\href@noop {}
  {\bibfield  {journal} {\bibinfo  {journal} {arXiv.org}\ } (\bibinfo {year}
  {2015})},\ \Eprint {http://arxiv.org/abs/1508.05882v2} {1508.05882v2}
  \BibitemShut {NoStop}%
\bibitem [{\citenamefont {Reed}\ \emph {et~al.}(1988)\citenamefont {Reed},
  \citenamefont {McCowan}, \citenamefont {Walsh}, \citenamefont {Delgado},\
  and\ \citenamefont {McColskey}}]{Indium:Reed:1988}%
  \BibitemOpen
  \bibfield  {author} {\bibinfo {author} {\bibfnamefont {R.~P.}\ \bibnamefont
  {Reed}}, \bibinfo {author} {\bibfnamefont {C.~N.}\ \bibnamefont {McCowan}},
  \bibinfo {author} {\bibfnamefont {R.~P.}\ \bibnamefont {Walsh}}, \bibinfo
  {author} {\bibfnamefont {L.~A.}\ \bibnamefont {Delgado}}, \ and\ \bibinfo
  {author} {\bibfnamefont {J.~D.}\ \bibnamefont {McColskey}},\ }\href {\doibase
  10.1016/0025-5416(88)90578-2} {\bibfield  {journal} {\bibinfo  {journal}
  {Mat. Sci. Eng. A-Struct.}\ }\textbf {\bibinfo {volume} {102}},\ \bibinfo
  {pages} {227} (\bibinfo {year} {1988})}\BibitemShut {NoStop}%
\bibitem [{\citenamefont {Khalil}\ \emph {et~al.}(2012)\citenamefont {Khalil},
  \citenamefont {Stoutimore}, \citenamefont {Wellstood},\ and\ \citenamefont
  {Osborn}}]{Khalil:2012jr}%
  \BibitemOpen
  \bibfield  {author} {\bibinfo {author} {\bibfnamefont {M.~S.}\ \bibnamefont
  {Khalil}}, \bibinfo {author} {\bibfnamefont {M.~J.~A.}\ \bibnamefont
  {Stoutimore}}, \bibinfo {author} {\bibfnamefont {F.~C.}\ \bibnamefont
  {Wellstood}}, \ and\ \bibinfo {author} {\bibfnamefont {K.~D.}\ \bibnamefont
  {Osborn}},\ }\href {\doibase 10.1063/1.3692073} {\bibfield  {journal}
  {\bibinfo  {journal} {J. Appl. Phys.}\ }\textbf {\bibinfo {volume} {111}},\
  \bibinfo {pages} {054510} (\bibinfo {year} {2012})}\BibitemShut {NoStop}%
\bibitem [{SUP()}]{SUPPLEMENT}%
  \BibitemOpen
  \href@noop {} {\bibinfo  {journal} {See supplementary material for machined cavity constructions, fabrication methods for
  micromachined cavities, and additional measurements of micromachined cavities
  with planar coupling}\ }\BibitemShut {NoStop}%
\bibitem [{\citenamefont {Huang}\ \emph {et~al.}(2010)\citenamefont {Huang},
  \citenamefont {Xu}, \citenamefont {Quan}, \citenamefont {Yuan},\ and\
  \citenamefont {Luo}}]{Inbump2}%
  \BibitemOpen
\bibfield  {journal} {  }\bibfield  {author} {\bibinfo {author} {\bibfnamefont
  {Q.}~\bibnamefont {Huang}}, \bibinfo {author} {\bibfnamefont
  {G.}~\bibnamefont {Xu}}, \bibinfo {author} {\bibfnamefont {G.}~\bibnamefont
  {Quan}}, \bibinfo {author} {\bibfnamefont {Y.}~\bibnamefont {Yuan}}, \ and\
  \bibinfo {author} {\bibfnamefont {L.}~\bibnamefont {Luo}},\ }\href {\doibase
  10.1088/1674-4926/31/11/116004} {\bibfield  {journal} {\bibinfo  {journal}
  {J. Semicond.}\ }\textbf {\bibinfo {volume} {31}},\ \bibinfo {pages} {116004}
  (\bibinfo {year} {2010})}\BibitemShut {NoStop}%
\bibitem [{\citenamefont {Broennimann}\ \emph {et~al.}(2006)\citenamefont
  {Broennimann}, \citenamefont {Glaus}, \citenamefont {Gobrecht}, \citenamefont
  {Heising}, \citenamefont {Horisberger}, \citenamefont {Horisberger},
  \citenamefont {K{\"a}stli}, \citenamefont {Lehmann}, \citenamefont {Rohe},\
  and\ \citenamefont {Streuli}}]{InBump4}%
  \BibitemOpen
  \bibfield  {author} {\bibinfo {author} {\bibfnamefont {C.}~\bibnamefont
  {Broennimann}}, \bibinfo {author} {\bibfnamefont {F.}~\bibnamefont {Glaus}},
  \bibinfo {author} {\bibfnamefont {J.}~\bibnamefont {Gobrecht}}, \bibinfo
  {author} {\bibfnamefont {S.}~\bibnamefont {Heising}}, \bibinfo {author}
  {\bibfnamefont {M.}~\bibnamefont {Horisberger}}, \bibinfo {author}
  {\bibfnamefont {R.}~\bibnamefont {Horisberger}}, \bibinfo {author}
  {\bibfnamefont {H.~C.}\ \bibnamefont {K{\"a}stli}}, \bibinfo {author}
  {\bibfnamefont {J.}~\bibnamefont {Lehmann}}, \bibinfo {author} {\bibfnamefont
  {T.}~\bibnamefont {Rohe}}, \ and\ \bibinfo {author} {\bibfnamefont
  {S.}~\bibnamefont {Streuli}},\ }\href {\doibase 10.1016/j.nima.2006.05.011}
  {\bibfield  {journal} {\bibinfo  {journal} {Nucl. Instrum. Meth. A}\ }\textbf
  {\bibinfo {volume} {565}},\ \bibinfo {pages} {303} (\bibinfo {year}
  {2006})}\BibitemShut {NoStop}%
\bibitem [{\citenamefont {Miller}\ and\ \citenamefont
  {Jhabvala}(2012)}]{Miller:2012vw}%
  \BibitemOpen
  \bibfield  {author} {\bibinfo {author} {\bibfnamefont {T.~M.}\ \bibnamefont
  {Miller}}\ and\ \bibinfo {author} {\bibfnamefont {C.~A.}\ \bibnamefont
  {Jhabvala}},\ }\href
  {http://ntrs.nasa.gov/archive/nasa/casi.ntrs.nasa.gov/20130013418.pdf}
  {\bibfield  {journal} {\bibinfo  {journal} {Proc. SPIE}\ }\textbf {\bibinfo
  {volume} {H}},\ \bibinfo {pages} {84532} (\bibinfo {year}
  {2012})}\BibitemShut {NoStop}%

\end{thebibliography}

\begin{thebibliography}{2}%
\makeatletter
\providecommand \@ifxundefined [1]{%
 \@ifx{#1\undefined}
}%
\providecommand \@ifnum [1]{%
 \ifnum #1\expandafter \@firstoftwo
 \else \expandafter \@secondoftwo
 \fi
}%
\providecommand \@ifx [1]{%
 \ifx #1\expandafter \@firstoftwo
 \else \expandafter \@secondoftwo
 \fi
}%
\providecommand \natexlab [1]{#1}%
\providecommand \enquote  [1]{``#1''}%
\providecommand \bibnamefont  [1]{#1}%
\providecommand \bibfnamefont [1]{#1}%
\providecommand \citenamefont [1]{#1}%
\providecommand \href@noop [0]{\@secondoftwo}%
\providecommand \href [0]{\begingroup \@sanitize@url \@href}%
\providecommand \@href[1]{\@@startlink{#1}\@@href}%
\providecommand \@@href[1]{\endgroup#1\@@endlink}%
\providecommand \@sanitize@url [0]{\catcode `\\12\catcode `\$12\catcode
  `\&12\catcode `\#12\catcode `\^12\catcode `\_12\catcode `\%12\relax}%
\providecommand \@@startlink[1]{}%
\providecommand \@@endlink[0]{}%
\providecommand \url  [0]{\begingroup\@sanitize@url \@url }%
\providecommand \@url [1]{\endgroup\@href {#1}{\urlprefix }}%
\providecommand \urlprefix  [0]{URL }%
\providecommand \Eprint [0]{\href }%
\providecommand \doibase [0]{http://dx.doi.org/}%
\providecommand \selectlanguage [0]{\@gobble}%
\providecommand \bibinfo  [0]{\@secondoftwo}%
\providecommand \bibfield  [0]{\@secondoftwo}%
\providecommand \translation [1]{[#1]}%
\providecommand \BibitemOpen [0]{}%
\providecommand \bibitemStop [0]{}%
\providecommand \bibitemNoStop [0]{.\EOS\space}%
\providecommand \EOS [0]{\spacefactor3000\relax}%
\providecommand \BibitemShut  [1]{\csname bibitem#1\endcsname}%
\let\auto@bib@innerbib\@empty

\bibitem [{\citenamefont {Collin}(1990)}]{Collin}%
  \BibitemOpen
  \bibfield  {author} {\bibinfo {author} {\bibfnamefont {R.~E.}\ \bibnamefont
  {Collin}},\ }\href@noop {} {\emph {\bibinfo {title} {{Field Theory of Guided
  Waves}}}},\ \bibinfo {edition} {2nd}\ ed.\ (\bibinfo  {publisher} {Wiley-IEEE
  Press, New York},\ \bibinfo {year} {1990})\BibitemShut {NoStop}%

\bibitem [{\citenamefont {Khalil}\ \emph {et~al.}(2012)\citenamefont {Khalil},
  \citenamefont {Stoutimore}, \citenamefont {Wellstood},\ and\ \citenamefont
  {Osborn}}]{Khalil:2012jr2}%
  \BibitemOpen
  \bibfield  {author} {\bibinfo {author} {\bibfnamefont {M.~S.}\ \bibnamefont
  {Khalil}}, \bibinfo {author} {\bibfnamefont {M.~J.~A.}\ \bibnamefont
  {Stoutimore}}, \bibinfo {author} {\bibfnamefont {F.~C.}\ \bibnamefont
  {Wellstood}}, \ and\ \bibinfo {author} {\bibfnamefont {K.~D.}\ \bibnamefont
  {Osborn}},\ }\href {\doibase 10.1063/1.3692073} {\bibfield  {journal}
  {\bibinfo  {journal} {J. Appl. Phys.}\ }\textbf {\bibinfo {volume} {111}},\
  \bibinfo {pages} {054510} (\bibinfo {year} {2012})}\BibitemShut {NoStop}%

\end{thebibliography}

%



\pagebreak
\onecolumngrid
\begin{center}
\textbf{\large Supplemental Materials: Demonstration of superconducting micromachined cavities}
\end{center}

\setcounter{equation}{0}
\setcounter{figure}{0}
\setcounter{table}{0}
\setcounter{page}{1}
\setcounter{section}{0}
\makeatletter
\renewcommand{\theequation}{S\arabic{equation}}
\renewcommand{\thefigure}{S\arabic{figure}}
\renewcommand{\bibnumfmt}[1]{[S#1]}
\renewcommand{\citenumfont}[1]{S#1}

\section{Descriptions of machined cavity constructions}
\begin{figure*}[h]
\begin{center}
\includegraphics[scale = 1,angle=0]{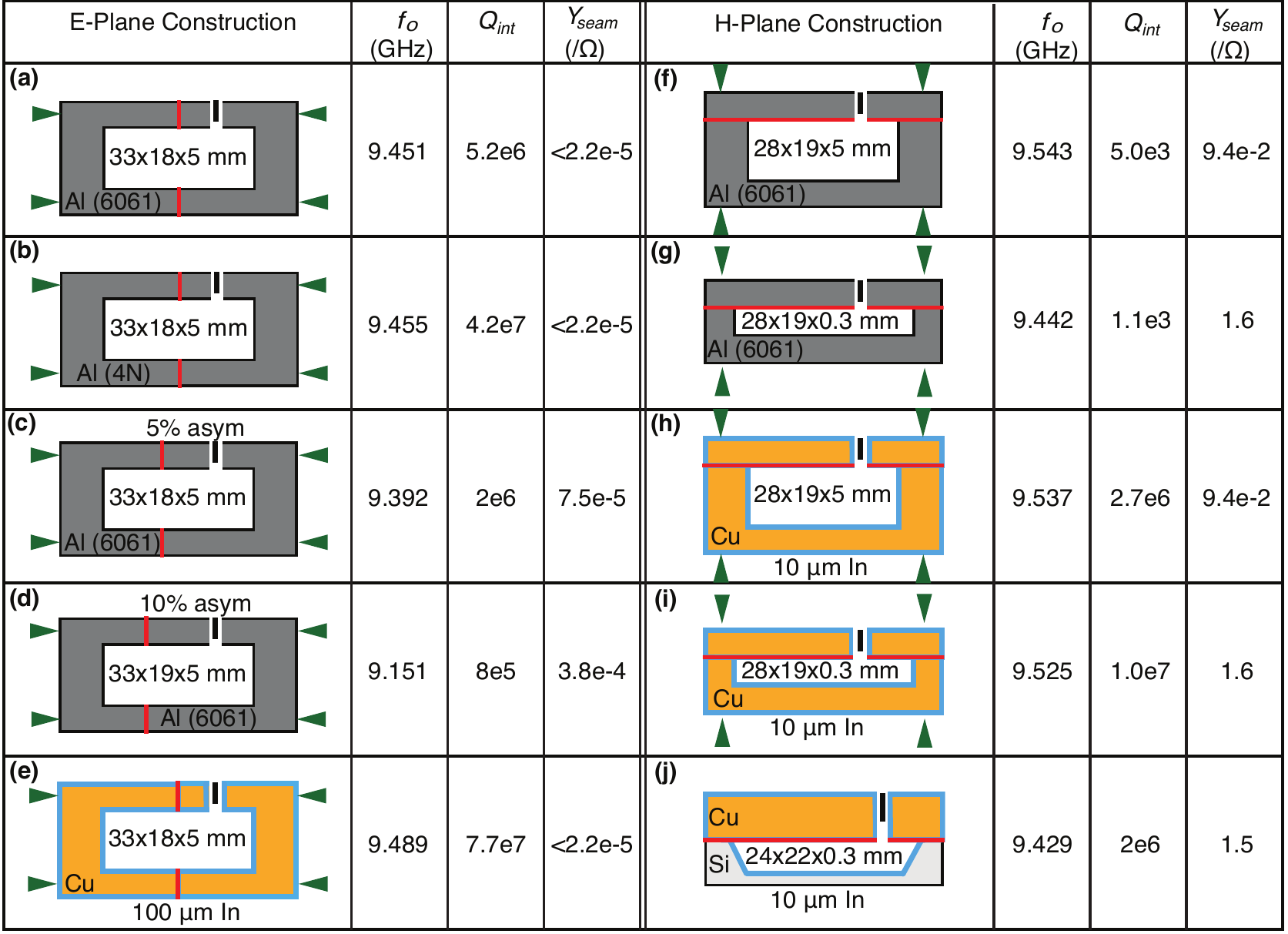}
\caption{
	\textbf{Assessing materials and seam losses for different 3D rectangular cavity constructions.}
    For each rectangular cavity, we show a T-plane cross-section of the construction, cold measured resonance frequency and internal Q, and $y_{seam}$. 
	}
\label{machinedcavitytable}
\end{center}
\end{figure*}

The table in figure \ref{machinedcavitytable} contains pictorial descriptions and measurement values for each data point presented in Fig. \ref{plot} of the main text. For each cavity construction, the admittance of the cavity presented to the seam is

\begin{equation}
Y_{seam}= \frac{L \int_{seam} |\vec{J_s} \times \hat{l}|^2 dl}
{\omega \mu_o \int_{tot} |\vec{H}|^2 dV}.
\end{equation}

\section{Fabrication of micromachined superconducting 3D resonators}

To create micromachined cavities, 1 mm thick $(100)$ silicon wafers are masked with lithographically patterned silicon nitride and etched by a KOH bath (30 \% KOH in water at 80 degrees C) to create 22 x 24 mm rectangular pits 300 $\mu$m deep.  The wet etch is anisotropic with selectivity to silicon's $(100):(111)$ planes, resulting in a rectangular recess with sidewalls determined by the crystal planes and RMS surface roughness of 20 nm.  The etch mask is removed and the wafer is metalized and bonded with another metalized substrate to create a sealed rectangular cavity.  The devices described here are coated with 6 nm titanium and 100 nm copper or gold which serves as a conducting layer onto which 10 $\mu$m of indium is electroplated, forming the cavity walls. For this process, an indium sulfamate plating bath solution purchased from Indium Corporation was operated at room temperature.  A rotating Pt-Ti mesh serves as the insoluble anode and the wafer serves as the cathode onto which indium precipitates from the bath.

For the devices measured in this work, an Instron 5969 compressive force testing machine is used to bond mating halves with indium contact areas of $50 - 500$ $\mathrm{mm}^2$ at forces between 4 and 40 kN.  Immediately prior to bonding, indium oxide is etched away with hydrochloric acid, but it is likely that 10 - 30 {\AA}  of oxide develops within minutes of exposure to air.

\section{Additional measurements of micromachined cavities with planar coupling}

The planar coupling design used in this work consisted of a circular aperture of omitted metalization positioned below a coplanar waveguide transmission line.  For radius $r<<\lambda$, the radiation of field $E_0$ through the aperture can be modeled as that of a dipole with $\vec{p}=\frac{2}{3}\epsilon_0 E_0 r^3$  oscillating at the cavity resonance frequency $\omega$.\cite{Collin} The coupling Q obeys the scaling $Q_C \propto h/\omega r^6$ provided that the width of the transmission line is comparable to the height $h$ above the aperture in its ground plane.

\begin{figure}[hb]
\includegraphics[scale = 1, angle = 0]{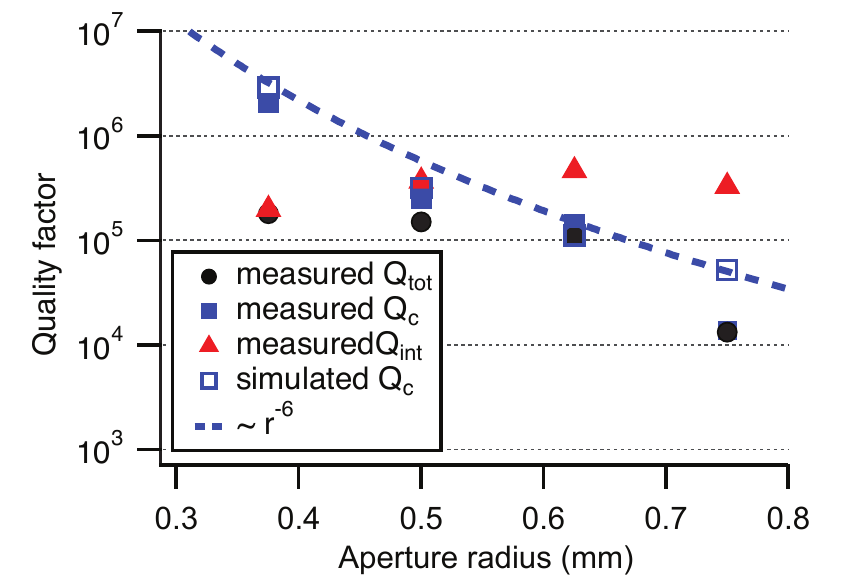}
\caption{
	\textbf{Quality factors of superconducting micromachined resonators with the described planar coupling scheme.}  
    All measurements were taken at temperatures $\approx$ 20 mK and the lineshape was fit\cite{Khalil:2012jr2} to obtain $Q_{tot}$, $Q_c$, and $Q_{int}$.  Simulated values are obtained by method of finite element analysis of electromagnetic fields using ANSYS HFSS.  The blue dashed line shows the expected scaling $Q_C \propto h/\omega r^6$.
    }
\label{SiPCBsummary}
\end{figure}

We conducted several measurements of micromachined resonators in the style described by Fig. \ref{umachine}. In Fig. \ref{SiPCBsummary}, we show results from four such devices made with different aperture radii. Measured coupling Qs compare well with simulation. Internal Qs measured here are less than 500,000, which we attribute to seam losses being the dominant source of dissipation.

\end{document}